\documentstyle{article}
\begin{document}
\title{\Large {\bf The hole spectral function and the relationship between
overlap functions, natural orbitals and the one-body density matrix in nuclei}
\footnote[1]{This work is partially supported by the
Bulgarian National
Science Foundation under the Contracts Nr.$\Phi$--32
and $\Phi$--406 and by the Royal Society and the Bulgarian
Academy of Sciences.}}
\author{A.N. Antonov, M.V. Stoitsov, M.K. Gaidarov, S.S. Dimitrova}
\date{\it Institute of Nuclear Research and Nuclear Energy,
Bulgarian Academy of Sciences, Sofia 1784, Bulgaria}
\maketitle
\vspace{-1.2cm}
\begin{center}
{\large and\\
 P.E. Hodgson\\
{\it Nuclear Physics Laboratory, Department of Physics, University of
Oxford, Oxford OX1-3RH, U.K.}}
\end{center}
\vspace{.3cm}

A method to calculate the hole spectral function in the discrete part of
the spectrum is suggested within the natural orbital representation of the
one-body density matrix of $A$-nucleon system using its relationship with the
overlap functions of the eigenstates in the $(A-1)$-nucleon system.
\vspace{1cm}\\
The cross-section of direct nucleon removal processes is determined by the
spectral function which contains the information on the nuclear structure and
is interpreted as the probability for the removal of a nucleon with given
momentum and energy from the target nucleus with $A$ nucleons (e.g., [1--25]).
In particular, in the plane-wave impulse approximation the cross-section for
the direct knock-out process is proportional to the diagonal element $S({\bf
k},{\bf k},E)\equiv S({\bf k},E)$ of the hole spectral function (or matrix)
in the momentum representation:
\begin{equation}
S({\bf k},{\bf k^{\prime} };E)=\left \langle \Psi_{0}\left| a^{+}({\bf
k^{\prime} })\delta (E+\widehat{H}-E_{A}^{0})a({\bf k})\right| \Psi_{0}
\right \rangle,
\label{1}
\end{equation}
where $| \Psi_{0}\rangle$ is the ground state wave function of the target
nucleus with $A$ nucleons, $a^{+}({\bf k^{\prime}})$ and $a({\bf k})$ are
creation and annihilation operators for a nucleon with momentum ${\bf
k^{\prime}}$ and
${\bf k}$, respectively, $\widehat{H}$ is the Hamiltonian of the system with
$(A-1)$-nucleons and $E_{A}^{0}$ is the ground state energy of the target
nucleus. If the latter has a total spin and parity $J^{\pi}=0^{+}$,
then introducing a complete set of eigenstates of $\widehat{H}$ for the system
of $(A-1)$-nucleons $| \Psi_{f}\rangle$ (where the state $| \Psi_{f}\rangle$ is
characterized by the energy $E_{f}$ with both discrete and continuous values
and by other discrete and continuous quantum numbers) the hole spectral
function can be written in the form:
\begin{eqnarray}
S({\bf k},{\bf k^{\prime} },E)&=&\sum_{f} \hspace{-.5cm} \int
\hspace{.2cm} \langle
\Psi_{0}\left| a^{+}({\bf k^{\prime}}) \right|  \Psi_{f}\rangle \langle
\Psi_{f} \left| a({\bf k}) \right| \Psi_{0} \rangle \delta(E+E_{f}-
E^{0}_{A})\\
&\equiv & \sum_{f} \hspace{-.5cm} \int \hspace{.2cm} \Phi_{f}^{*}({\bf
k^{\prime}}) \Phi_{f}({\bf k}) \delta(E+E_{f}-E_{A}^{0}),
\end{eqnarray}
where
\begin{equation}
\Phi_{f}({\bf k}) \equiv \langle \Psi_{f} \left| a({\bf k}) \right|
\Psi_{0} \rangle
\label{4}
\end{equation}
is the overlap function in the momentum representation [26--28].

The methods used to calculate the spectral function are reviewed, e.g. in
\cite{7,14,16}. The use of the independent-particle shell model (when the
overlap function (4) is equal to the single-particle wave function of the
occupied state) cannot explain the fragmentation or spreading of the hole
strength. This is because, due to the residual
interaction, the hole state in the target nucleus is not an eigenstate of the
$(A-1)$-nucleon system and its strength is distributed over several states
of the final system.

In this work we suggest a method to calculate the hole spectral functions using
essentially the one-body density matrix (OBDM) of $A$-nucleon system in the
natural orbital representation \cite{29} and its relationships with the natural
orbitals (which diagonalize the OBDM) and the overlap functions (4). An
expansion of the latter in the basis of the natural orbitals is used. The
following two reasons can justify the use of the method:

1) Recently the diagonalization of the realistic one-body density matrix of the
correlated nuclear ground state obtained by various correlation methods
\cite{16}, such as the Jastrow method [30--32], as well as the generator
coordinate method \cite{16,33,34} and the coherent density fluctuation model
\cite{16,34,35} gave reliable information on the natural orbitals and
occupation numbers in nuclei. These quantities correspond to the realistic
behaviour of nuclear characteristics which are sensitive to the short-range
nucleon-nucleon correlations, such as the nucleon and cluster momentum
distributions, the mean kinetic and removal energies, radii and others. The
natural orbitals in nuclei, as well as those in other fermion systems, such as
$^{3}He$ liquid drops \cite{36}, are strongly localized and quite different
from the overlap functions and from the mean-field type orbitals
\cite{30,34,36,37}. Thus, it is of importance to apply the natural orbitals
corresponding to realistic OBDM obtained in correlation theoretical methods to
calculate the hole spectral function $S({\bf k},{\bf k^{\prime}},E)$.

2) The basic quantity which is necessary to calculate the spectral
function (3) is the overlap function (4). We show in this paper
that the hole spectral function in the discrete part  of the spectrum can be
calculated by using the general relationship \cite{37} which connects the
asymptotic behaviour of the one-body density matrix with the overlap functions
of the $(A-1)$-particle system eigenstates. This relationship is of
general importance because it enables one to obtain quantities connected with
the bound eigenstates of the $(A-1)$-particle system (such as overlap
functions, spectroscopic factors and separation energies) by means of the exact
OBDM (or by a realistic one obtained in a given correlation method) of the
ground state of the $A$-particle system. In this way, the hole spectral
function in the discrete part of the spectrum can be, in principle, calculated
on the basis of the OBDM of the $A$-particle system.

Now we introduce the necessary quantities which are used in the
theoretical method to calculate the hole spectral function.

The one-body density matrix of the ground state $| \Psi_{0}\rangle$
of the $A$-nucleon system has the form
\begin{equation}
\rho (x,x^{\prime})=\langle \Psi_{0} \left| a^{+}(x)a(x^{\prime} ) \right|
\Psi_{0} \rangle ,
\label{5}
\end{equation}
where $x\equiv \{{\bf r}\sigma \tau \}$ labels spatial, spin and isospin
coordinates and $a^{+}(x)$,$a(x^{\prime})$ are the creation and annihilation
operators.

The natural orbitals (NO) $\varphi_{a}(x)$ are defined \cite{29} as the
complete orthonormal set of single-particle wave functions which diagonalize
the OBDM:
\begin{equation}
\rho (x,x^{\prime} )=\sum_{a} N_{a} \varphi_{a}^{*}(x) \varphi_{a}
(x^{\prime}).
\label{6}
\end{equation}
The eigenvalues $N_{a}$ ($0\leq N_{a} \leq 1$, $\displaystyle \sum_{a}
N_{a}=A$) are the natural occupation numbers. We note that the sum (6)
is over the discrete states determined by the finite-range NO
 $\varphi_{a}({\bf k})$.

The OBDM (5) can be presented also in the form:
\begin{equation}
\rho (x,x^{\prime} )=\sum_{f} \hspace{-.5cm} \int \hspace{.2cm}
\Phi_{f}^{*}(x) \Phi_{f}(x^{\prime}),
\label{7}
\end{equation}
where $\Phi_{f}(x)=\langle \Psi_{f} \left| a(x) \right| \Psi_{0} \rangle $
is the overlap function in the coordinate representation.

The overlap functions can be expanded in the basis of the natural orbitals
(e.g., in momentum space):
\begin{equation}
\Phi_{f}({\bf k})=\sum_{a} \langle \varphi_{a} | \Phi_{f} \rangle
\varphi_{a}({\bf k})
\label{8}
\end{equation}
The hole spectral function is then given by the expression:
\begin{equation}
\begin{array}{lcl} \displaystyle
S({\bf k},{\bf k^{\prime}},E)&=&\displaystyle\sum_{a,b} \varphi_{a}^{*}({\bf
k^{\prime} }) \varphi_{b}({\bf k}) \sum_{f} \hspace{-.5cm} \int \hspace{.2cm}
\langle \Phi_{f} | \varphi_{a} \rangle \langle \varphi_{b} | \Phi_{f} \rangle
\delta(E+E_{f}-E_{A}^{0})\\
&\displaystyle\equiv & \displaystyle\sum_{a,b} \varphi_{a}^{*}
({\bf k^{\prime}}) \varphi_{b}({\bf k})S_{ab}(E),
\end{array}
\label{9}
\end{equation}
where
\begin{equation}
S_{ab}(E)\equiv \sum_{f} \hspace{-.5cm} \int \hspace{.2cm} \langle \Phi_{f} |
\varphi_{a} \rangle \langle \varphi_{b} | \Phi_{f} \rangle
\delta(E+E_{f}-E_{A}^{0}).
\label{10}
\end{equation}
The quantity (for which different notations exist, e.g. \cite{14,15}):
\begin{equation}
\theta_{a,f}\equiv S_{a,f}^{1/2}\equiv \langle \varphi_{a} | \Phi_{f}
\rangle
\label{11}
\end{equation}
from (8) and (9) is the amplitude of the contribution of the orbital $a$ to the
overlap function for the eigenstate $| \Psi_{f}\rangle$. We mention that the
quantity (11) determines both the spectroscopic factor of the state
$| \Psi_{f}\rangle$ \cite{28}
\begin{equation}
S_{f}^{A-1}\equiv \langle \Phi_{f} | \Phi_{f} \rangle =\sum_{a}{\left|
\theta_{a,f} \right|^{2}}=\sum_{a} S_{a,f}=\sum_{a}{\left| \langle \varphi_{a}
| \Phi_{f} \rangle \right|^{2}}
\label{12}
\end{equation}
and the occupation probability of the orbital $a$:
\begin{equation}
N_{a}=\sum_{f} \hspace{-.5cm} \int \hspace{.2cm} \left| \theta_{a,f}
\right|^{2}=\sum_{f} \hspace{-.5cm} \int
\hspace{.2cm} S_{a,f}=\sum_{f} \hspace{-.5cm} \int \hspace{.2cm} \left|
\langle \varphi_{a} | \Phi_{f} \rangle \right|^{2}.
\label{13}
\end{equation}
In general, for a given orbital $a$, only a limited subset of states $f$ of the
residual nucleus contribute to the sums (10) and (13).

The function $S_{ab}(E)$ (given by Eq.(10) and often called
also "spectral function") can be
rewritten \cite{5} introducing the different states $| \Psi_{f}\rangle$ of the
residual nucleus: i) the bound states $| \Psi_{E_{\nu},\alpha}\rangle$ with
energy $E_{\nu}$ and degeneracy quantum number $\alpha$, and ii) the continuum
states $| \Psi_{E_{f},c}\rangle$ with energy $E_{f}$ and the channel index $c$
which specifies the channel where there is an incoming wave (all other channels
contain only outgoing waves), as well as all degeneracies like spin
projections etc. Then Eq.(10) becomes:
\begin{equation}
\begin{array}{lcl} \displaystyle
S_{ab}(E)&=&\displaystyle\sum_{\nu,\alpha} \langle
\Phi_{\nu \alpha}| \varphi_{a}\rangle
\langle \varphi_{b}| \Phi_{\nu \alpha} \rangle
\delta(E+E_{\nu}-E_{A}^{0})\\
&\displaystyle +&\displaystyle\sum_{c} \langle
\Phi_{E_{f}=E_{A}^{0}-E,c}|
\varphi_{a} \rangle \langle
\varphi_{b}| \Phi_{E_{f}=E_{A}^{0}-E,c} \rangle \theta
(E_{A}^{0}-E_{A-1}^{thr.}-E)\\
&\displaystyle \equiv &\displaystyle S_{ab}^{d.s.}(E)+S_{ab}^{c.s.}(E),
\end{array}
\label{14}
\end{equation}
where $\Phi_{\nu \alpha}$ and $\Phi_{E_{f},c}$ are the overlap functions
associated with the bound and continuum eigenstates of the residual nucleus and
$E_{A-1}^{thr.}$ is the threshold for particle decay of this nucleus. If the
latter is a nucleon threshold, then $E_{A-1}^{thr.}=E_{A-2}^{0}$, where
$E_{A-2}^{0}$ is the ground-state energy of the nucleus with $A-2$ nucleons
\cite{5}. The hole-spectral function (14) contains two parts: i) the spectral
function in the discrete part of the spectrum $S_{ab}^{d.s.}(E)$, and ii) the
spectral function in the continuum of the hole spectrum $S_{ab}^{c.s.}(E)$ with
$E\leq E_{A}^{0}-E_{A-2}^{0}$.

As can be seen from Eqs.(9) and (14) the hole spectral function is essentially
connected with the natural orbitals $\{\varphi_{a}\}$ and the overlap
functions $\Phi_{f}$ and their relationship with the OBDM. Now we shall
outline briefly this relationship.

In the case of spherical symmetry the OBDM can be written in the form:
\begin{equation}
\rho(x,x^{\prime} )=\sum_{qlj} \rho^{(qlj)}(r,r^{\prime} )\sum_{m}
Y_{ljm}^{*}(\Omega,\sigma)Y_{ljm}(\Omega^{\prime} , \sigma^{\prime} ),
\label{15}
\end{equation}
where the radial part of the OBDM is:
\begin{equation}
\rho^{(qlj)}(r,r^{\prime} )=\sum_{f} \hspace{-.5cm} \int \hspace{.2cm}
\Phi_{f}^{(qlj)}(r) \Phi_{f}^{(qlj)}(r^{\prime} ).
\label{16}
\end{equation}
In the above equations $\Phi_{f}^{(qlj)}(r)$ is the radial part of the overlap
function, $Y_{ljm}(\Omega,\sigma)$ is the spin-angular function, $q$ denotes
the nature (proton and neutron) of the overlap function and $l$, $j$ are
angular and total momentum quantum numbers.

It is known \cite{28} that the overlap functions associated with the bound
states of the $(A-1)$- and $(A+1)$-nucleon systems are eigenstates of a
single-particle Schr\"{o}dinger equation in which the mass operator plays the
role of a potential. Due to the finite range of the mass operator, the
asymptotic behaviour of the radial part of the neutron overlap functions for
bound states $\nu$ (labeled by $\nu$=0, 1, ... with increasing energy) of
the $(A-1)$-nucleon system is given by \cite{26,27,37}:
\begin{equation}
\Phi_{\nu}^{(qlj)}(r)\rightarrow C_{\nu}^{(qlj)}exp(-k_{\nu}^{(qlj)}r)/r,
\label{17}
\end{equation}
where
\begin{equation}
k_{\nu}^{(qlj)}=\frac{1}{\hbar} [2m_{q}(E_{\nu}^{(qlj)}-E_{A}^{0})]^{1/2}.
\label{18}
\end{equation}
For protons some mathematical complications arise due to an additional
long-range part originating from the Coulomb interaction \cite{27}, though
everything from the neutron case remains valid. It is assumed in \cite{37} that
Eq.(17) is also valid for the overlap functions corresponding to the $(A-1)$
continuum.

The asymptotic form of the overlap functions (Eqs.(17) and (18)) determines the
asymptotic behaviour of the radial part of the OBDM
$\rho^{(qlj)}(r,r^{\prime} )$ \cite{37}. It is shown in \cite{37} that at large
values of $r^{\prime} \equiv a \rightarrow \infty $ one can derive the lowest
bound state overlap function by means of the radial part of the OBDM:
\begin{equation}
\Phi_{0}^{(qlj)}(r)=\frac{\rho^{(qlj)}(r,a)}{C_{0}^{(qlj)}exp(-k_{0}^{(qlj)}a)
/a},
\label{19}
\end{equation}
as well as the separation energy
\begin{equation}
\varepsilon_{0}^{(qlj)}=\hbar ^{2}k_{0}^{(qlj)2}/2m_{q}
\label{20}
\end{equation}
and the spectroscopic factor
\begin{equation}
S_{0}^{(qlj)}=\langle \Phi_{0}^{(qlj)}| \Phi_{0}^{(qlj)}\rangle.
\label{21}
\end{equation}
The normalization coefficient $C_{0}^{(qlj)}$ can be obtained from the
asymptotic form of the diagonal part of the radial OBDM:
\begin{equation}
\rho^{(qlj)}(a,a)\rightarrow
\left|C_{0}^{(qlj)}\right|^{2}exp(-2k_{0}^{(qlj)}a)/a^{2}.
\label{22}
\end{equation}
As shown in \cite{37}, the overlap functions for all bound states of the
$(A-1)$-nucleon system can be constructed from the OBDM repeating the above
procedure. For instance, the overlap function for the next bound state is:
\begin{equation}
\Phi_{1}^{(qlj)}(r)=\frac{\rho^{(qlj)}(r,a)-\Phi_{0}^{(qlj)}(r)
\Phi_{0}^{(qlj)}(a)}{C_{1}^{(qlj)}exp(-k_{1}^{(qlj)}a)/a}.
\label{23}
\end{equation}

In the case of the continuum contributions to the OBDM one can calculate the
particular sum over the scattering channels $c$: $\displaystyle \sum_{c}
[\Phi_{c}^{(qlj)}(r,E)C_{c}(E)]$, but not the overlap function for each
channel \cite{37}.

The method for calculating of the hole spectral function in the discrete part
of the spectrum (for non-degenerate  states $\nu$):
\begin{equation}
S^{d.s.}({\bf k},{\bf k^{\prime}},E)=\sum_{a,b} \varphi_{a}^{*}
({\bf k^{\prime}})
\varphi_{b}({\bf k})\sum_{\nu} \langle \Phi_{\nu}| \varphi_{a} \rangle
\langle \varphi_{b}| \Phi_{\nu} \rangle \delta(E+E_{\nu}-E_{A}^{0})
\label{24}
\end{equation}
from a given theoretical correlation method consists in the following
procedure:\\
1) By diagonalizing the one-body density matrix of the $A$-particle system
ground state one obtains the natural orbitals $\{\varphi_{a}({\bf
k})\}$ (e.g., as in [30--35,16]);
2) The bound-state overlap functions $\Phi_{\nu}$ and
separation energies $\varepsilon_{\nu}$ are calculated on the basis of the
one-body density matrix following the approximate method described above
(Eqs.(19)-(23)). 3) The amplitudes of the contribution
of the natural orbital $a$ to the overlap function $\langle \varphi_{a}|
\Phi_{\nu} \rangle $ are calculated and the results substituted in Eq.(24).
The overlap functions of the discrete states indeed allow the
corresponding part of the spectral function to be calculated directly
from Eq.(3). However, the third step of the procedure also makes it
possible to calculate simultaneously on the same footing the
amplitudes (11) and some particular terms of the sums in Eqs.(12),
(13) and (24). The spectroscopic amplitudes (11) can then be calculated
with shell model single-particle wave functions and compared with
those obtained from natural orbitals.
It can be seen from Eqs.(1) and (5) that the energy integral of the hole
spectral function (1) defines the one-body density matrix in the momentum
representation
\begin{equation}
\int_{-\infty}^{E_{F}^{-}} dE S({\bf k},{\bf k^{\prime}},E)=\rho ({\bf k},
{\bf k^{\prime}}),
\label{25}
\end{equation}
where $E_{F}^{-}$ is a negative quantity whose absolute value is equal to the
separation energy of the $A$-nucleon system \cite{28}.

We emphasize that the method described above enables one to obtain the hole
spectral function in the discrete part of the spectrum (i.e. the integrand of
the left-hand side of Eq.(25) in the energy interval between
$E_{A}^{0}-E_{A-2}^{0}$ and $E_{F}^{-}$) on the basis of the one-body density
matrix calculated in a given correlation method. The knowledge of $\rho
({\bf k},{\bf k^{\prime}})$ can give some information on the remaining
part of the integrand in the left-hand side of Eq.(25), namely the hole
spectral function in the continuum part of the spectrum in the energy
interval between $-\infty $ and $E_{A}^{0}-E_{A-2}^{0}$.

In this paper we suggest a new theoretical method to obtain the hole spectral
function in the discrete part of the spectrum. The method is based on the
natural orbital representation in nuclear theory and uses essentially both the
natural orbitals and overlap functions as well as their relationship with the
OBDM. Thus the theoretical point of the method consists in the possibility of
using the OBDM which is related to the properties of the $A$-nucleon system
to calculate the hole spectral function which determines the cross-section of
the nucleon removal processes and gives information on the structure of the
$(A-1)$-nucleon system. The applications of the method can serve also as a
test of the predictions of the correlated methods concerning the OBDM of the
correlated ground state of the $A$-nucleon system.

Our program to apply the suggested method includes two stages: i) calculations
of overlap functions on the basis of realistic OBDM from a given correlation
method and studies of their properties, and ii) calculations of hole spectral
functions in the discrete part of the spectrum and comparison with available
experimental data. The results from the fulfilment of this program will be
given elsewhere.

\end{document}